\documentclass[aps,prd,nofootinbib,square,sort&compress,10pt,fleqn,showpacs,longbibliography]{revtex4-1}
\usepackage{amssymb}
\usepackage{textcomp}
\usepackage[fleqn]{amsmath}
\usepackage{empheq}
\usepackage[utf8,latin1]{inputenc} 
\usepackage[T1]{fontenc}
\usepackage{pslatex}
\usepackage{csquotes}
\usepackage[x11names]{xcolor}
\usepackage{empheq}
\usepackage{xcolor}
\definecolor{rkka}{RGB}{219,66,32}
\usepackage{color}
\allowdisplaybreaks
\usepackage{mathtools}
\usepackage{eucal}
\usepackage{graphicx}
\usepackage{float}
\usepackage[pdfencoding=auto,psdextra]{hyperref}
\usepackage{bookmark}
\usepackage{url}
\usepackage{breakurl}
\usepackage{epsfig}
\usepackage{bm}
\usepackage[
english, % Englische Rechtschreibung
ngerman  % neue Deutsche Rechtschreibung (wenn hier auch Überschriften)
]{babel} % Silbentrennung etc.
\usepackage{marvosym} % Sonderzeichen
\usepackage{emerald}
\usepackage{slantsc}
\usepackage{array}

\let\tempcal \mathcal
\usepackage{dutchcal}

\let\mathcal\tempcal

\makeatletter
\newcommand{\mathleft}{\@fleqntrue\@mathmargin0pt}
\newcommand{\mathcenter}{\@fleqnfalse}
\makeatother

\def\be{\mathleft\begin{equation}}
\def\ee{\end{equation}}
\def\ba{\mathleft\begin{eqnarray}}
\def\ea{\end{eqnarray}}
\def\Asu{\begin{subequations}}
\def\esu{\end{subequations}}

\def\a{\alpha}

\def\d{\delta}

\def\l{\lambda}
\def\m{\mu}
\def\n{\nu}
\def\la{\label}

\def\le{\left}
\def\ri{\right}

\begin{document}
%%%%%%%%%%%%%%%%%%%%%%%%%%%%%%%%%%%%%%%%%%%%%%%%%%%%%%%%%%%%%%%%%%%%
\title{The Science of Fundamental Catalogs}
\author{Sergei M. Kopeikin}
\affiliation{Department of Physics \& Astronomy, University of Missouri, 322 Physics Bldg., Columbia, Missouri 65211, USA}
\email{E-mail: kopeikins@missouri.edu}
\author{Valeri V. Makarov}
\affiliation{US Naval Observatory, Astrometry Department, 3450 Massachusetts Ave., NW Washington DC 20392, USA}
\email{E-mail: valeri.makarov@navy.mil}
\begin{abstract}
This review paper discusses the science of astrometric catalogs, their current applications and future prospects for making progress in fundamental astronomy, astrophysics and gravitational physics. We discuss the concept of fundamental catalogs, their practical realizations, and future prospects. Particular attention is paid to the astrophysical implementations of the catalogs such as the measurement of the Oort constants, the secular aberration and parallax, and asteroseismology. We also consider the use of the fundamental catalogs in gravitational physics for testing general theory of relativity and detection of ultra-long gravitational waves of cosmological origin.  
\end{abstract}
%\keywords{}
\maketitle
\tableofcontents
%%%%%%%%%%%%%%%%%%%%%%%%%%%%%%%%%%%%%%%%%%%%%%%%%%%%%%%%%%%%%%%%%%%%%%%%
\section{Introduction}
Fundamental astronomy is currently an integral part of modern gravitational physics and astrophysics, which rely upon two pillars -- astrometry and celestial mechanics. The defining task of fundamental astronomy is to build the inertial celestial reference frame, which is used to determine coordinates,
velocities, and accelerations of astronomical bodies and to predict their past, present, and future dynamical evolution in the course of time. Reference objects of fundamental astronomy are 
stars and quasars, which are benchmarks materializing the inertial reference frame on the sky. Quasars are located at large distances from the Solar System so that their trigonometric parallaxes are difficult or even impossible to measure even now. Historically, the inability to detect the annual parallax was used as an argument against the 
Copernicus heliocentric
theory, but it was based on misapprehension of astronomical distance scales. Parallaxes of just a few dozen stars could 
be measured until the early 20th century, when specialized
astrograph telescopes and photographic plates facilitated a breakthrough in precise astrometry.  Celestial coordinates and 
velocities are traditionally determined as angles on the celestial sphere and their time derivatives (proper motions),
in contrast to the Cartesian three-dimensional coordinates and velocities. For a long time, astrometric observations of stars and planets were interpreted in the framework of spherical astronomy and the Newtonian celestial mechanics in the Euclidean space. 

The situation had changed about 100 years ago with the advent of a new generation of large optical telescopes and the emergent understanding of the true nature of spiral nebulae and the size of the universe, accompanied by the advent of special and general theory of relativity and flourishing of astrophysics \citep{Trimble_1995}. Fundamental astronomy has undergone dramatic technological changes and explored a broader range of electromagnetic spectrum radiation
operating from the ground and from space.
A cardinal improvement in the precision of astrometric measurements of positions and parallaxes of stars and quasars 
has been achieved (in some cases approaching the level of $\simeq$10 $\mu$as) by making use of the Very Long Baseline Interferometry (VLBI) \citep{fomalont_2002evn,fomalont_2004NewAR,Sanna_2017} and the Gaia space satellite \citep{Castelvecchi_2016Natur}. 

One should notice that the measurement of time is an integral part of fundamental astronomy as time is one of the four coordinates of spacetime manifold, which is the arena where all the physical and astronomical phenomena take place. Precise
time determination is a cornerstone discipline and technology on its own but its synergy with fundamental astronomy
has never been more essential than today. Modern technology allows one to manufacture atomic clocks with a relative fractional instability of the order of a few parts in $10^{-18}$ \citep{Poli_2013NCimR,Lisdat_2018RPPh}. A global network of such most stable and accurate clocks produces a practical realization of the International  Atomic Time (TAI) scale at the level of $10^{-16}$ \citep{Petit_2010HiA}, which is the basic time scale of planetary and lunar ephemerides \citep{kopeikin_2011book}, binary star orbital dynamics \citep{Anglada_2006}, exoplanetary search \citep{Kaplan_2005,pexo_2019ApJS}, pulsar timing astronomy \citep{Lorimer_2012book}, etc. 
   
Technological achievements in high-precision astronomical measurements of coordinates and velocities of celestial bodies along with manufacturing of ultra-accurate quantum clocks open new fascinating research opportunities in the field of fundamental astronomy and its astrophysical applications which are briefly reviewed in the present article. 
   
\section{Astrometry of Fundamental Catalogs}

\subsection{The concept of a fundamental catalog}

A fixed inertial system of coordinates on the sky is called the celestial reference system. A practical realization of the celestial reference system is called a celestial reference frame (CRF) as defined by a list of reference objects forming a fundamental catalog \citep{Kopff_1936MNRAS}. Fundamental catalogs are based on absolute measurements of the reference objects without using previous information about their positions and motions. The concepts
\enquote{absolute} and \enquote{fundamental} underwent subtle transformations over the past few decades. Differential or
relative astrometry, which is the counterpart of absolute, is a more understandable notion, applied mostly to
wide-angle instruments and surveys where astrometric determinations are performed in reference to sources
whose apparent places at the time of observation are assumed to be known with superior accuracy from a
different catalog. What defines an astrometric measurement as absolute? In the pre-Hipparcos era, when ground-based
meridian circles were the driving force of absolute astrometry, a measurement was considered as being independent
of other catalogs
if it was performed relative to a \enquote{fixed}, or at least long-term stable, direction with respect to the surface of
Earth such as two artificial light sources in long evacuated tubes or a plumb line realized through a liquid mercury reference horizon.
Being attached to a constantly rotating and wobbling Earth, these reference directions could only be assumed
to be fixed if the instantaneous orientation parameters that is the components of the angular velocity vector, were known to superior accuracy. The Earth orientation parameters as well as the orbit of Earth around the Sun, were determined by special optical measurements
referred to the fundamental astrometric stars using a different set of instruments and techniques. This mutual
dependence of Earth orientation and absolute astrometry of stars may look like a vicious circle, but there was no viable
alternative until the advent of the {\it Very-Long Baseline Interferometry} with linked radio telescopes, and with
it, of {\it global} astrometric techniques. The term \enquote{global} is sometimes used as replacement to \enquote{absolute},
which is not exact, because the truly global space astrometry with Hipparcos and Gaia, as discussed in the following, is not fully absolute \citep{vallenari_2018FrASS}.

Until the discovery of quasars by M. Schmidt in 1963, the main reference objects of the CRF were stars. A fundamental catalog contains a limited number of standard stars whose coordinates and proper motions are measured with utmost astrometric precision and considered as known. The catalog is used for measuring relative positions and velocities of other astronomical objects. The main plane of the CRF is the celestial equator at a fixed standard epoch and one 
of the coordinate axes is placed in the direction of vernal equinox, that is, the point of intersection of the celestial equator with the ecliptic. The origin of the coordinate system is implicitly placed at the center of mass
of the Solar System. Since the Earth moves in space along its orbit around the Sun and its axis of rotation undergoes precession, the specific realization of the celestial reference frame depends on the standard epoch of the equinox and on our knowledge of the constant of precession \citep{Sovers_2000}. This brings in the ephemerides of the Newtonian theory of gravity to the realization of the inertial frame in the form of ephemerides of Earth and 
major planets of the Solar System, as well as an elaborate model of Earth rotation \citep{kovalevsky_2012}. 
Through this choice of coordinates, Earth orientation parameters and theory, precise timing, and absolute
astrometry are the three inseparable parts of the CRF.

The first fundamental catalog of the northern sky (abbreviated as FC) was compiled by A. Auwers and published in 1879 \citep{auwers_1879}. The FC catalog included positions and proper motions of 539 fundamental stars. The second fundamental catalog (NFK) consisting of 925 stars was published in 1907 by J. Peters \citep{peters_1907VeABD}. It extended the FC from the northern to southern sky. The third fundamental catalog (FK3) was produced by A. Kopff in 1938 \citep{Kopff_1938AN,Kopff_1939AN}. Progressive development of the star-based fundamental catalog resulted in the compilation of FK4 and further FK5 catalogs by W. Fricke \citep{FK4_1963fcf,FK5_1988VeARI}. FK5, published in 1988, included 1535 fundamental stars from FK4 as a primary standard along with 3117 stars as an extension. All the fundamental catalogs up to FK5 used observations made with ground-based optical instruments \citep{Fricke_1985CeM,perryman_2012EPJH}. On the other hand, the Sixth Catalog of Fundamental Stars (FK6) is the result
of the combination of the Hipparcos satellite data given in the Hipparcos Catalog (ESA 1997) with the ground-based data listed in the basic part of the FK5-Part I \citep{Wielen_1999VeARI}. 

\subsection{International Celestial Reference Frame (ICRF)}
A major disadvantage of the stellar fundamental catalogs is the proper motion of the defining stars. Stars move in the
tangential plane because of the combined effects of the Galactic rotation and their own peculiar velocity
due to the distribution of Galactic orbits. Because of the non-random, or systemic, components of stellar
proper motions, the celestial reference frame has a residual rotation and, strictly speaking, cannot be considered as inertial. Furthermore, a fraction of nearby stars comprise
moving groups and kinematic associations with complex, non-random patterns on the sky \citep{2001ASPC..244...57M}. 
Even when taken
as a whole and averaged, a sample of stars does not provide a fixed direction required to achieve absolute astrometry.
Distant galaxies and active galactic nuclei (AGN) are much further away from the Solar System and their true proper motions 
are believed to be negligibly small. They are much better celestial reference objects to build an inertial reference 
frame on the sky. However, galaxies have noticeable angular sizes preventing precise measurement of their photometric centers  in the optical wavelengths,  which 
reduces the accuracy of the catalog. This problem is much less essential for quasars, 
which are luminous nuclei of the most active and distant galaxies. Therefore, the International Astronomical Union (IAU) decided to build the International Celestial 
Reference Frame (ICRF) by making use of quasars as fundamental reference objects. Millions of quasars have been
cataloged using various methods in the optical and near-infrared domain, but only several thousand are  bright 
enough in the radio to be observed by the VLBI, which is a self-sufficient and global technique inherently linked to the
Earth orientation parameters and precision timing, thus closely approaching the concept of absoluteness.

The first realization of the ICRF1 was completed in 1998 and contained 212 defining quasars \citep{ICRF1_1998AJ}. It was adjusted to FK5, in order to preserve continuity with the series of the stellar fundamental catalogs, and delivered an order of magnitude improvement in the realization of the inertial reference frame on the sky. A decade later, the ICRF2 catalog consisting of 295 reference quasars, was released \citep{ICRF2_2009ITN}. It was subsequently replaced by the ICRF3 \citep{icrf3_2020A&A} as adopted by the IAU during its XXX-th General Assembly in 2018 \citep{ICRF3_2018evn}. The ICRF3 contains positions of 4536 quasars as measured at frequencies 8.4 and 2.3 GHz (X$/$S band). 
Among them 303 sources are identified as defining sources. These are rather uniformly distributed on the sky 
and determine the axes of the frame. The quasar positions at 8.4$/$2.3 GHz are supplemented with positions of 824 sources at 24 GHz (K band), and with position of 678 sources at 32$/$8.4 GHz (Ka$/$X band). 
The positions of ICRF3 sources are adjusted to those in the ICRF2 at the epoch J2015.0 with an accuracy of 30 $\mu$as. The advanced feature of the ICRF3 is the accounting for the galactocentric acceleration of the Solar System, which causes a secular aberration effect \citep{kopmak_2006AJ} of 5.8 $\mu$as/yr in the apparent proper motions of quasars \citep{MacMillan_2019AA}.

The attempt to introduce alternative catalogs obtained by different VLBI configurations at higher electromagnetic frequencies
is related to the problem of frequency-dependent astrometric position of interferometric cores located along
parsec-scaled jets of AGNs \citep{2008AAKov} which may reach $\sim 1$ mas, thus, compromising the formal
precision of ICRF coordinates. 
The effect of frequency-dependent astrometric core positions in ICRF3 measured by the phase and group delay interferometric technique has been studied more thoroughly by \citet{porcas_2009}, who argued that if the core shift, $\Delta x$, is described by the frequency power-law, $\Delta x\sim\nu^\beta$, with the index $\beta=1$, then, there is no difference between the phase and group delay positions of the core and the absolute radio astrometry gives the position of the AGN jet base without appreciable error. Recently, a more comprehensive investigation of this problem has been conducted \citep{kovalev_2019MNRAS}. It revealed that the offsets between the core positions is frequency-dependent and may reach nearly 0.5 mas at 2 and 8 GHz. 
It can also change in the course of time with a characteristic variability of the individual core positions of about 0.3 mas. 

The practical realization of ICRF3 heavily relies upon the advanced astronomical and geophysical modeling of VLBI observations \citep{icrf3_2020A&A}. The modeling has been gradually improved over the past 30 years resulting in the latest exhaustive version of the VLBI data processing algorithm documented in the Technical Note 36 of the International Earth Rotation and Reference Systems Service (IERS) \citep{IERS_2010conventions}. It includes a comprehensive description of the earth's rotation parameters along with its 
general-relativistic theory of orbital motion in the Solar System, propagation of radio waves in the ionosphere and 
troposphere, and a self-consistent post-Newtonian theory of reference coordinate systems and gravitational Shapiro's time delay \citep{kopeikin_2011book,Soffel_2017JGeod}. New IERS conventions are currently in preparation and are to be released in 2022. 

\subsection{Hipparcos}

A great boost to the development of the optical fundamental catalog was given by the launch of the Hipparcos 
astrometric satellite in 1989 \textcolor{blue}{\mbox{[\url{https://www.cosmos.esa.int/web/hipparcos}]}}. After $\sim3.5$ 
years of the generally successful mission, the main Hipparcos catalog and several ancillary data sets were published 
in 1997 \citep{Hipparcos_1997A&A}. It contains 118,218 entries (single and multiple stars) distributed over the entire sky with a mean density of approximately 3 stars per square degree. The Hipparcos catalog was adjusted to the ICRF1 through a combination of radio star positions and by a few additional indirect techniques. Combining the Hipparcos
5-parameter astrometry with older ground-based observations lead to the creation of the sixth fundamental catalog (FK6) published in two parts: FK6(I) in 1999 \citep{FK6_1999VeAR}, and FK6(III) 2000 \citep{FK6-iii_2000}. The FK6 contains several solutions for the defining stars. The classical one is the single-star mode solution (SI mode) with a mean error in proper motion of 590 $\mu$as/yr. The other two solutions are called the long-term prediction (LTP) and short-term prediction (STP) modes. They have been introduced in response to the quasi-instantaneous nature of the Hipparcos measurements which could not discriminate between a binary star with orbital period of a few years and a single star. The LTP and STP modes are the most precise solutions for single Hipparcos stars with a typical mean error of proper motion of 930 $\mu$as/yr.   

The Hipparcos technology revolutionized the field of astrometry and galactic astronomy \citep{perryman_1998PhT,
perryman_2012aaa}. It significantly extended the reference frame and achieved a much higher precision of standard astrometric solutions for a single star based on five model parameters -- right ascension $\alpha$, declination $\delta$, parallax $\varpi$, and two components of proper motion in the plane of the sky, $\mu_\alpha$ and $\mu_\delta$ \citep{kovalevsky_2012}. Combining the Hipparcos catalog with the sixth  observable parameter - the radial velocity of the star - measured independently by means of stellar spectroscopy, effectively allowed one to start studying kinematics of stars in the six-dimensional phase space. This provides a complete description of the trajectories of stars in the neighborhood of the Solar System that is vitally important for an adequate understanding of the processes of formation and dynamical evolution of the Milky Way \citep{perryman_2012aaa}. 

\subsection{Gaia}

The Gaia astrometric satellite \citep{2016A&A...595A...1G}, launched on 19 December 2013 \textcolor{blue}{\mbox{[\url{https://www.cosmos.esa.int/web/gaia/launch}]}}, is the next step forward after the Hipparcos. The astrometric objectives of the mission are to collect and build a 3-dimensional catalog of more than 1 billion stars and 500,000 quasars and to discover thousands of new asteroids and Jupiter-size exoplanets. Like the Hipparcos mission, Gaia measures positions, proper motions, and parallaxes of stars, which (along with precise ground-based and Gaia's own spectroscopic measurements of radial velocities) provide the most comprehensive census of  about 1\% of the total stellar population of our Galaxy. The Gaia catalog is released in several steps. As of today, two releases of the Gaia catalog -- DR1 \textcolor{blue}{\mbox{[\url{https://www.cosmos.esa.int/web/gaia/dr1}]}}, 
\citep{2016A&A...595A...2G} and DR2 \textcolor{blue}{\mbox{[\url{https://www.cosmos.esa.int/web/gaia/dr2}]}} -- have been published.

The Gaia DR1 catalog includes about 1.14 billion sources. Its five-parameter astrometric solution is given for about 2 million stars from the intersection with the Tycho-2 catalog \citep{tycho2_2000}
which were treated as single stars without taking their radial velocity into account. The Gaia-CRF1 reference frame is aligned with the ICRF2 radio catalog at the 0.1 mas level at epoch J2015.0, and is non-rotating with respect to the ICRF2 to within 0.03 mas/yr \citep{DR1_ICRF}. On the other hand, it was found that the Hipparcos reference frame has a residual rotation with respect to the Gaia-DR1 frame at a rate of 0.24 mas/yr \citep{DR1_astrometry}. 

The DR2 extends the CRF to a larger number of sources amounting to 1.69 billion stars with the mean epoch J2015.5. The DR2 five-parameter astrometric solution  significantly extends that of the DR1, and includes about 1.33 billion stars.  Moreover, the DR2 provides radial velocities for more than 7.2 million stars with stellar magnitudes in the range $4\leq m\leq 13$ and an effective temperature in the range $3550\leq T\leq 6900$ K. This yields a full six-parameter astrometric  characterization for those stars. The reference frame of the DR2 is a densification of the Gaia-CRF1 from 2191 frame-fixing sources to 556869 sources uniformly distributed across almost the entire sky except along the Galactic equator \citep{Gaia-CRF2}. The densification of the quasar grid improved the determination and removal of the residual global rotation of the celestial frame caused by the observed proper motion of the frame-fixing sources. The densification does not change the orientation of the Gaia-CRF2 coordinate axes which is still defined by a much smaller number of radio-loud quasars with accurate positions in the ICRF3 catalog. 

At the time when this review was in preparation, the Gaia consortium has announced a third data release (EDR3) which includes a few catalogs and documentation, available online at the Gaia website \textcolor{blue}{[\url{https://www.cosmos.esa.int/web/gaia/early-data-release-3}]}. The contents and survey properties of EDR3 are summarized in \citep{gaia-edr3}. The source list of EDR3 is mostly the same as that for Gaia DR2 but it does feature new additions and there are some notable changes. The creation of the source list for Gaia EDR3 includes enhancements that make it more robust with respect to high proper motion stars and the disturbing effects of spurious and partially resolved sources. Briefly, EDR3 includes astrophotometric solution for approximately 1.8 billion sources brighter than m$>$21. For 1.5 billion of those sources, parallaxes, proper motions, and the G$_{\rm BP}$-G$_{\rm RP}$ color are also available. EDR3 also includes 7 million radial velocities from Gaia DR2 after removal of a small number of spurious values. EDR3 represents an updated materialization of the celestial reference system -- the Gaia-CRF3 -- which is based solely on extra-galactic sources in the optical bandwidth and ICRF3 positions. For bright stars (G < 13 mag), however, the Hipparcos catalog was used to correct for a residual spin of the proper motion field. The internal consistency of the EDR3 CRF realization may also suffer from the split of the source list into two categories, viz., 5-parameter solutions, and somewhat less reliable 6-parameter solutions.

Gaia EDR3 represents a significant advance over Gaia DR2, with parallax precision increased by 30 percent, proper motion precision increased by a factor of 2, and astrometric systematic errors suppressed by 30--40 percent for the parallaxes and by a factor ~2.5 for the proper motions. The photometry also benefits from higher precision with much better homogeneity across color, magnitude, and celestial position of the sources. Consistent definitions of the G, G$_{\rm BP}$, and G$_{\rm RP}$ passbands are achieved over the entire magnitude and color range with no systematic above the 1 percent level. The Gaia team works on the data release DR3 which will be based on the improved version of Gaia-EDR3. 

\subsection{Other Catalogs and Databases}
It is worth mentioning that besides the fundamental catalogs there exists a large number of other astronomical catalogs and databases of various astronomical objects grouped together by a common type, morphology, origin, or method of discovery. Wikipedia lists all existing astronomical catalogs on this website \textcolor{blue}{[\url{https://en.wikipedia.org/wiki/List_of_astronomical_catalogues}]}. Among these, the Tycho-2 catalog of the 2.5 million brightest stars \citep{tycho2_2000} is one of the most useful and frequently used in fundamental astronomy. It contains resolved binary systems with separations down to 0.8 arcsec \citep{2002A&A...384..180F} and proper motions with accuracy up to 2.5 mas/yr. It was amended a few years ago with the data taken from Gaia-DR1 resulting in the new 5-parameter Tycho-Gaia astrometric solution (TGAS), reaching a positional accuracy of stars of 0.3 mas in positions of epoch and 1 mas/yr in proper motions \citep{tgas_2015A}. 

The VizieR database should also be mentioned, which groups in an homogeneous way thousands of astronomical catalogs gathered for decades by the Centre de Donn\'ees de Strasbourg (CDS) and participating institutes \citep{VizieR}. This web-accessible database has developed into a powerful tool enabling researchers and users to retrieve and combine astronomical information across various disciplines and domains, including fundamental astrometry.

\section{Astrophysics of Fundamental Catalogs}
\subsection{The Oort constants}
Astrophysical applications of fundamental catalogs are ubiquitous. Exact positions, trigonometric parallaxes, and proper motions of stars from the catalog along with their radial velocities (when available) form the observational basis to study stellar kinematics of the Milky Way, globular clusters and dwarf spheroidal galaxies from the Local Group. Stellar kinematics of the stars near the Sun allows us to measure the important Oort's constants abbreviated as $A$, $B$, 
$C$, and $K$ in the linearized two-dimensional model of rotation of the galactic disc neglecting its thickness and represented by close orbits of stars. Two of  the constants, $A$ and $B$, which characterize the azimuthal shear and 
vorticity of the velocity field, were introduced by Oort himself \citep{Oort_1927_constants} for the idealized axisymmetric model of Milky Way's rotation. The two other constants, currently denoted as $C$ and $K$, characterize the radial shear and divergence of the velocity field and account for a non-axisymmetric component of the distribution of stellar velocities \citep{ogorodnikov_1932,milne_1935MNRAS} \footnote{The constant $K$ is denoted as $B'$ in Ogorodnikov's paper \citep{ogorodnikov_1932}. The letter $K$ for the forth Oort's constant respects Ogorodnikov's notation of the gravitational force vector ${\bm K}=(K_x,K_y)=(K,0)$ according to Eq. (6) in \citep{ogorodnikov_1932}.}. The four constants define the local rotational properties of the kinematic vector field of stellar velocities and give us insight in the mass density profile of the galactic disc within the solar neighborhood ($A$ and $B$ constants) as well as allows us to evaluate the physically important parameters of the Galactic spiral density wave ($C$ and $K$ constants) \citep{bobbaj_2013,bobylev_2020AstL}.  

Measuring the Oort constants has been a high priority task for fundamental astronomy since 1927 when the Oort's theory of Galactic rotation was formulated \citep{Oort_1927_constants} and the astrometric measurements of proper motions of stars became sufficiently accurate. Nonetheless, before the Hipparcos mission, the numerical estimates of the Oort constants had a large dispersion depending on the origin of data and methods used to compile different catalogs \citep{kerrlynden1986,olling1998,makarov_2007AJ}. One of the reasons is that prior to the Hipparcos breakthrough, the astrometric measurements of stellar proper motions relied on fundamental catalogs like FK4 and FK5, whose reference
systems were tied to the orbits of Earth and major planets and, thus,  were sensitive to the precession of the vernal equinox with respect to the ecliptic. Satellite measurements of proper motions, on the
other hand, are  tied, directly or indirectly, to quasars and distant galaxies and are not sensitive to the possible residual rotation of the reference frame of the fundamental catalog. Additional systematic errors in the values of the Oort constants appeared in previous studies due to the axisymmetric idealization of the Milky Way with almost circular orbits of stars, mathematically represented by a linear tensor. With the growing accuracy and larger samples of stars in astrometric catalogs, these systematic errors became noticeable, so that more accurate models of the motion of the Solar System with respect to the Local Standard of Rest (LSR) had to be worked out along with the necessity to account for the vertical component of the systemic motion of stars and the mode mixing 
effect \cite{olling_2003ApJ}. A comprehensive study of the Oort constants using astrometric measurements for 304267 main-sequence stars within a typical distance of 230 pc, taken from the Gaia-DR1 TGAS catalog, was conducted by \citet{bovy_2017MNRAS}, who obtained for the local velocity field generated by closed orbits:
\ba\la{1}
A&=&+15.3\pm 0.4\;\; \mbox{km s$^{-1}$ kpc$^{-1}$}\;,\qquad\mbox{(azimuthal shear)}\\\la{2}
B&=&-11.9\pm 0.4\;\; \mbox{km s$^{-1}$ kpc$^{-1}$}\;,\qquad\mbox{(vorticity)}\\\la{3}
C&=&-{~}3.2\pm 0.4\;\; \mbox{km s$^{-1}$ kpc$^{-1}$}\;\,,\qquad\mbox{(radial shear)}\\\la{4}
K&=&-{~}3.3\pm 0.6\;\; \mbox{km s$^{-1}$ kpc$^{-1}$}\;,\qquad\mbox{(divergence)}
\ea
with no color trend over a wide range of stellar populations. The measured values of the Oort constants \eqref{1}--\eqref{4} correspond to the Sun's  peculiar velocity $u_0 = 8.99 \pm 0.31$ km/s, $w_0 = 7.22 \pm 0.22$ km/s, while $10\leq v_0\leq 22$ km/s is sensitive to the stellar population of reference,
because of the asymmetric drift \footnote{See \citep[\S 24.3]{intromodastr_book} and plots of $(u_0,v_0,w)$ in the entry [26] on website     
\textcolor{blue}{\mbox{[\url{https://gist.github.com/jobovy/b002421ca865c93991f08f0eda4e72de}]}} and  for more detail.}. Comparison with previous determination of the Oort constants and the solar peculiar velocity made by \citet{makarov_2007AJ} on the sample of 42339 non-binary Hipparcos stars with
accurate parallaxes, using a vector spherical harmonic formalism, shows a good agreement with \eqref{1}--\eqref{4} within the statistical error except of the $v_0$ component of the solar peculiar velocity which is strongly affected by the asymmetric drift.

Will the linearized Oort and Ogorodnikov-Milne models remain adequate in the post-Gaia future$\mathord{?}$ Perhaps, a
three-dimensional mathematical representation of the Galactic velocity field with the origin at the Galactic center and
a larger number of fitting parameters will become necessary at some point. The vector spherical harmonic formalism
is still intrinsically two-dimensional, and therefore, not well suited to capture the increasingly available radial dimension of the
field. A galactocentric model based on a vector generalization of solid spherical harmonics, or other combinations of
orthogonal vector functions may provide inroads in the formal part of the problem. For a well-conditioned,
three-dimensional  Galactocentric model, the kinematics of stars beyond the Galactic center has to be accurately
known. This goal will require pushing
the distance limits of the fundamental catalog to \enquote{the other half} of the Galaxy. Interstellar extinction of light 
within the Galactic disk is the main obstacle,
and exploring space astrometry in the infrared, as proposed in the Gaia-NIR successor mission \citep{2019arXiv190712535H}, 
will open up new fascinating possibilities in this area of research.
  
\subsection{The secular aberration}
While the Solar System moves with respect to the LSR with the peculiar velocity $u_0,v_0,w_0$, the LSR itself moves around the center of the Milky Way along a circular path, which is characterized by the local parameters of Galactic rotation: $R_0$ -- the radial distance to the center of the Milky Way, $\Theta_0$ and $(d\Theta/dR)_{R_0}$ -- the linear velocity of the circular orbital motion and its radial gradient, and $\omega_0=\Theta_0/R_0$. In the axisymmetric approximation, these parameters are directly related to the Oort constants $A$ and $B$, that is, $\omega_0=A-B$ and $(d\Theta/dR)_{R_0}=-(A+B)$. The IAU recommended value for the LSR rotational velocity $\Theta_0= 220$ km/s and the distance to the galactic center $R_0= 8.5$ kpc. These values do not directly match the measured values of the Oort constants \eqref{1}, \eqref{2} due to the fact that in the solar neighborhood, non-circular streaming motions are important. Indeed, comparison of \eqref{3}, \eqref{4} with \eqref{1}, \eqref{2} shows that $|C|\simeq |K|\simeq 0.2 |A|$ and they cannot be ignored in the determinations of the circular velocity of the LSR from the local kinematics of stars. Since the parameters of Galactic rotation are used for evaluating the total mass distribution inside the Milky Way and testing general relativity with binary pulsars \citep{wex2014,weisberg_2016ApJ}, their measurement being done independently of the Oort constants plays a crucial role \citep{fomalont_2004NewAR,Reid_2014ApJ,veracol_2020}. 

It was noticed that the centrifugal acceleration $a=d\Theta_0/dt$ of the circular motion of the LSR is expressed in terms of the product $a=\omega_0\Theta_0$ which can be measured from the analysis of a systematic vector field of the apparent proper motions of distant quasars over the entire sky induced by the the effect of the secular aberration of light. The amplitude of the secular aberration effect was determined at 5.8 $\mu$as/yr \citep{MacMillan_2019AA}. It is large enough
to be taken into account in the ICRF3 catalog in certain high-accuracy applications \citep{Truebenbach_2017}. The deviation from this nominal value caused by the peculiar motion of the Solar System with respect to the LSR is less than 0.8 $\mu$as/yr \citep{kopmak_2006AJ,titov_2011A&A,titov_2018A&A,MacMillan_2019AA}, which is sufficiently negligible. 

The measurement of the secular aberration is a good and independent quality test of the Gaia astrometric solution. The recently announced astrometric solution based on Gaia EDR3 yields the direction of the secular acceleration towards  the point on the celestial sphere with the coordinates $\alpha=269.1\deg\pm 5.4\deg$, $\delta=-31.6\deg\pm 4.1\deg$, with a value corresponding to a proper motion amplitude of $5.05\pm 0.35$  $\mu$as$\cdot$yr$^{-1}$ \citep{klioner_2020aber}. These results are in good agreement with the acceleration vector of the Solar System expected from the current models of the Galactic gravitational potential. It is anticipated that future Gaia data releases will provide the estimates of the secular aberration effect with uncertainties substantially below 0.1 $\mu$as$\cdot$yr$^{-1}$.

At the current level of precision of astrometric observations, the secular aberration alone cannot be used for independent measurement of the Galactic rotation parameters. This will be achieved in future astrometric missions (e.g., STARE, NEAT, THEIA \citep{janson_2018haex,theia_2017arXiv} or Gaia-NIR \citep{gaianir_2018IAUS}, etc.) that will allow position and parallax measurements to better than 1 $\mu$as and
proper motions to better than 1 $\mu$as/yr. Nonetheless, a recent paper by \citet{bovy_2020} demonstrates that measurement of the secular aberration effect in combination with relative accelerations obtained from binary pulsar orbital decays \citep{Chakrabarti_2020} allows one to determine all of the parameters describing the dynamics of our local Galactic environment, including the circular velocity of the Sun, and its
derivative, the local angular frequency, the Oort constants, and the Sun's motion with respect to the LSR.

\subsection{The secular parallax}  

The Solar System moves with respect to the center of the Milky Way, which itself traverses with respect to the cosmological reference frame defined by the isotropy of the Cosmic Microwave Background (CMB) \citep{weinberg_2008book}. The overall velocity of the motion of the Solar System barycenter with respect to the CMB is directed towards the galactic coordinates $(l, b) = (263.99^\circ\pm 0.14^\circ, 48.26^\circ\pm 0.03^\circ)$ and has a magnitude of $v_\odot=369\pm 0.9$ km/s \citep{planck_dipole_2014}. Distant cosmological objects (galaxies, quasars) participate in the Hubble expansion and move fast in the radial direction with respect to the Milky Way. Nonetheless, they also have peculiar velocities with respect to the CMB. The peculiar motion of the Milky Way and distant galaxies is caused by the inhomogeneities of the large scale structure of the universe. Several researchers have recently suggested that catalog's analysis of the proper motions of distant cosmological objects can provide meaningful consistency tests of the standard model of cosmology \footnote{The number count of the galaxies can be useful tracers of the cosmological parameters in addition to the proper motions \citep{Maartens_2018,bengaly_2018JCAP,pant_2019JCAP}.}, place independent constraints on the numerical value of the Hubble constant and the linear growth rate of cosmic structures, and be
instrumental in the search for the nature of dark matter \citep{hall_2019MNRAS-cos,Paine_2020}.
 
As an example, let us consider how the measurement of the vector field of galactic proper motions can constrain the Hubble constant $H_0$. 
The proper motion of each galaxy consists of several components. The first one is caused by the velocity of the Solar System 
 (i.e., observer), ${\bm v}_\odot$, with respect to the CMB, which is called the secular parallax effect \citep{kardashev_1986SvA}. This effect manifests as a dipole harmonic in the proper motion field of the galactic population in the direction collinear to ${\bm v}_\odot$ and amounting to $78 \;d^{-1}\sin\beta$ $\mu$as/yr Mpc, where $\beta$ is the angle between the galaxy and ${\bm v}_\odot$. The second component of the proper motion is caused by the peculiar velocity of the galaxy, ${\bm v}_g$, with respect to the CMB. The peculiar velocities ${\bm v}_g$ of sufficiently close galaxies are not distributed randomly in space since they are caused by the large scale structure inhomogeneities of the local supercluster \citep{hall_2019MNRAS-cos}. The proper motion caused by ${\bm v}_g$ has approximately the same magnitude and distance dependence as the secular parallax proper motion. The third component of the proper motion is caused by the secular aberration. The secular aberration does not depend on the distance. Moreover, it can be measured independently by observing solely the proper motion of very distant quasars \citep{kopmak_2006AJ,titov_2011A&A,titov_2018A&A} and after that subtracted. 

The fact that the peculiar velocities of galaxies are strongly correlated with the large scale structure allows us to compute the best statistical estimates of the corresponding proper motions and subtract them from the relative proper motion to get the secular parallax effect for each galaxy   \citep{hall_2019MNRAS-cos}. The evaluation of the secular parallax makes it possible
 to measure the proper distance $d$ to the galaxy independently of the astronomical \enquote{distance ladder}  based on standard candles like cepheids, etc. At the same time, a spectroscopic measurement of the redshift $z$ of the galaxy yields the quantity $z=H_0d$. Making use of the distance $d$ from the secular parallax, independent constraints on the value of the Hubble constant $H_0$
 can be obtained. This method was used by \citet{Paine_2020} who estimated a first preliminary limit of 3500 $\mu$as/yr Mpc on the secular parallax amplitude using proper motions of 232 nearby galaxies from Gaia-DR2. This limit can be improved by a factor of 10 at the end of Gaia mission to set a tighter constraint on the Hubble parameter.

\subsection{Asteroseismology and parallax zero-point}

A completely new application of the data from fundamental catalogs emerged from their growing capability to deliver a consistency check of the theory of the internal structure of stars. Conventionally, it was broad-band photometry and trigonometric parallax, $\varpi$, of stars that have been used to determine their physical characteristics: the effective surface temperature $T_*$, radius $R_*$, and luminosity $L_*$, under the condition that extinction $A_\l$ and reddening have been measured as well \citep{intromodastr_book}. The photometric parallax of a star is given by 
\ba\label{ghe56}
\Pi&=&c_\l\le(\frac{R_{\rm bb}}{R_\odot}\ri)^{-1}\le(\frac{T_*}{T_\odot}\ri)^{-2}\;
\ea
where
\ba 
c_\l&=&10^{0.2\le(m_\l+BC_\l+5-A_\l-M_{{\rm bol}\odot}\ri)}\;,
\ea
$m_\l$ and $BC_\l$ are the star's apparent magnitude and bolometric correction at a given wavelength, $M_{{\rm bol}\odot}=+4.75$ is the Sun's absolute magnitude, and $R_{\rm bb}$ is the radius of the black body having the effective temperature $T_*$. Notice that, generally speaking, $R_{\rm bb}$ is close but not equal to $R_*$.

Asteroseismology provides the theoretical relation between the star's radius and the global seismic parameters \citep{khan_2019A&A}
\ba \label{jki45}
%\le(\frac{M}{M_\odot}\ri)&=&\le(\frac{\nu_{\rm max}}{\nu_{{\rm max}\odot}}\ri)^3\le(\frac{\Delta\nu}{\Delta\nu_\odot}\ri)^{-4}\le(\frac{T_*}{T_\odot}\ri)^{3/2}\;,\\
\le(\frac{R_*}{R_\odot}\ri)&=&\le(\frac{\nu_{\rm max}}{\nu_{{\rm max}\odot}}\ri)\le(\frac{\Delta\nu}{\Delta\nu_\odot}\ri)^{-2}\le(\frac{T_*}{T_\odot}\ri)^{1/2}\;,
\ea
where $\Delta\nu$ is the average frequency separation between the modes of stellar oscillations and $\nu_{\rm max}$ is the frequency corresponding to the maximum observed oscillation power,  both normalized to the solar values of the helioseismological parameters \citep{Kjeldsen_1995}. Making use of \eqref{jki45} in \eqref{ghe56} yields an asteroseismic parallax
\ba
\varpi_{\rm sei}&=&{\sf A}\,c_{\l}\le(\frac{\nu_{\rm max}}{\nu_{{\rm max}\odot}}\ri)^{-1}\le(\frac{\Delta\nu}{\Delta\nu_\odot}\ri)^{2}\le(\frac{T_*}{T_\odot}\ri)^{-5/2}\;,
\ea
where ${\sf A}\simeq 1$ is the scaling factor taking into account the deviation of $R_{\rm bb}$ from $R_*$ and normalizing the global asterosesmological parameters $\Delta\nu$ and $\nu_{\rm max}$ to the solar values $\Delta\nu_\odot$ and $\nu_{{\rm max}\odot}$ \citep{Kjeldsen_1995,khan_2019A&A}. 

The astreroseismic parallax $\varpi_{\rm sei}$ can be independently computed from corresponding observations of stars and compared with the trigonometric parallax $\pi$ of the same stars  from a fundamental astrometric catalog like Hipparcos or Gaia-DR2.  This way, the scaling factor ${\sf A}$ can be determined along with the so-called global parallax zero-point $\pi_0$ of the astrometric catalog \citep{lindegren_1995A&A,makarov_1998A&A,butkevich_2017A&A}. The value of ${\sf A}\simeq 1$ is used to calibrate the theory of stellar oscillations while the knowledge of $\pi_0\simeq 0$  provides the correct value of trigonometric parallaxes. The parallax zero-point can be also determined independently by making use of distant quasars which have negligible trigonometric parallaxes for $z>0.1$,  but a difficulty arises with possible
dependence of the zero-point on magnitude, which affects a large number of brighter stars. 

A comparison of the different types of parallaxes has been recently conducted by \citet{khan_2019A&A} who used the Kepler mission data \textcolor{blue}{\mbox{[\url{https://www.nasa.gov/mission_pages/kepler/main/index.html}]}} to compare the results with Gaia-DR2. They find that there is no absolute standard within asteroseismology, because different seismic approaches to the parallax determination problem extensively produce a parallax zero-point, which is fairly different from the Gaia-DR2 $\pi_0\simeq -29$ $\mu$as determined by \citet{DR2_2018A&A} by making use of 555934 sources identified as quasars.  
Similar analysis was done by \citet{zinn_2019ApJ} who used asteroseismic data of evolved red giant branch stars from the APOKASC-2 catalog to determine the zero-point offset of Gaia DR2 parallaxes. Testing asteroseismology with Gaia DR2 has been also undertaken by \citet{hall_2019MNRAS} who used a sample of 5576 red clump stars from the Kepler mission field.

Broadly speaking, the parallax zero-point $\pi_0$ is the systematic parallax error which applies to all objects in a fundamental catalog and, therefore, profoundly affects many important applications, including the fundamental distance scale.  For Gaia, it comes from specific thermally-induced variations of the basic angle with time, which are periodic functions of the Sun's aspect angle \citep{butkevich_2017A&A}. The Gaia basic angle monitoring system (BAM) is not sufficient to track this variation at the required level of accuracy, resulting in a complicated instrument calibration scheme. It is not easy to estimate the parallax zero point because the error may be both magnitude and color-dependent through the functional dependencies of the calibration models and intrinsic poor conditioning. Some estimates of the parallax zero-point in EDR3 catalog are presented by \citet{gaia-edr3_2020arXiv201203380L} along with a complete list of references to previous papers and in-depth discussion of the problem. Notably, compared with
Gaia-DR2, the astrometric calibration models in EDR3 have been extended, and the spin-related distortion model includes a self-consistent determination of
basic-angle variations, improving the global parallax zero point. The global parallax zero point of EDR3 is about -17 $\mu$as \citep{gaia-edr3_2020arXiv201201742L}. The parallax zero point issue will be hopefully resolved, at the
level of a few $\mu$as, in future data releases that will rely upon further improved calibration models.

\section{Gravitational Physics of Fundamental Catalogs}
\subsection{Testing general relativity}
The most precise test of general relativity before the advent of gravitational wave astronomy was 
achieved with timing of compact binary pulsars \citep{krawex_2009,wex2014}. The compact binary pulsars have short orbital periods and move fairly fast ($v/c\sim 10^{-3}$), which makes them ideal celestial laboratories for measuring a number of various relativistic parameters. Most of the parameters are associated with the relativistic effects in the conservative (1-st and 2-nd) post-Newtonian approximations. Gravitational waves carry out the orbital energy and angular momentum of the source binary system causing the orbital period, $P_b$, to decay. The rate of the decay was computed in general relativity (GR) with different mathematical techniques (see review by \citet{Damour_1987book} for historical details) and is given by \citep{petersmath_PhysRev,peters_PhysRev,Damour_1983grr,k85,schaefer_1985AnPhy}
\ba\label{pb12} 
\le(\frac{\dot P_b}{P_b}\ri)^{\rm GR}&=&-\frac{96\pi}{5}\frac{G^{5/3}}{c^5}\frac{m_1m_2}{\le(m_1+m_2\ri)^{1/3}}\le(\frac{P_b}{2\pi}\ri)^{-8/3}\le(1+\frac{73}{24}e^2+\frac{37}{96}e^4\ri)
\le(1-e^2\ri)^{-7/2}\;,
\ea
where $m_1$ and $m_2$ are the masses of the pulsar and its companion, $e$ is the eccentricity of the binary system, $G$ is the universal gravitational constant, and $c$ is the fundamental speed of gravity (equal to the speed of light in vacuum).

The observed value of $\dot P^{\rm obs}_b$ is contaminated by a number of relativistic effects caused by the galactic motion of the binary pulsar with respect to the Solar System \citep{dt_1991ApJ}
\ba\label{pb23}
\le(\frac{\dot P_b}{P_b}\ri)^{\rm obs}&=&\le(\frac{\dot P_b}{P_b}\ri)^{\rm GR}+\le(\frac{\dot P_b}{P_b}\ri)^{\rm gal}\;,
\ea
where the numerical value of $\dot P^{\rm gal}_b$ crucially depends on our knowledge of the stellar kinematics of the disk of the Milky Way. According to \citet{dt_1991ApJ} 
\ba\label{pb34}
\le(\frac{\dot P_b}{P_b}\ri)^{\rm gal}&=&-\frac{\Theta_0}{c}\le[\cos l+\frac{\d-\cos l}{\d-\cos 2l}\le(1-\frac{A+B}{A-B}\sqrt{\d-\cos 2l}\ri)\ri]\;,
\ea
where $\Theta_0$ is the circular velocity of the LSR, $l$ is the galactic longitude of the binary pulsar, $\d=d/R_0$, $d$ is the distance between the binary pulsar and the Sun, $R_0$ is the distance from the LSR to the center of the Milky Way, $A$ and $B$ are the Oort constants. 

Testing general relativity in binary pulsars is based on measurement of the five classic (Keplerian) parameters of the orbit along with any two (post-Keplerian) relativistic parameters of the 1-st post-Newtonian approximation of general relativity \citep{dt_1992PhRvD}. Usually, the two post-Keplerian parameters are -- the secular drift of periastron $<\dot\omega>=f_{\dot\omega}(m_1,m_2,e,P_b)$, and the Doppler+Einstein gravitational delay $\gamma=f_{\gamma}(m_1,m_2,e,P_b)$ -- which depend on the masses of the pulsar and its companion, orbital eccentricity and the orbital period. The eccentricity $e$ and the orbital period $P_b$ are known from the measured Keplerian parameters. Hence, measuring $<\dot\omega>$ and $\gamma$ allows to determine the masses $m_1$ and $m_2$. These values can be substituted to the right hand-side of \eqref{pb12} along with $P_b$ and $e$  yielding  $\le(\dot P_b/P_b\ri)^{\rm GR}_{\rm pred}$, which can be compared with the observed value $\le(\dot P_b/P_b\ri)^{\rm GR}$. 

Unfortunately, the direct comparison is impossible, as one can see from \eqref{pb23}, because of the galactic contribution $\le(\dot P_b/P_b\ri)^{\rm gal}$ to the observed orbital decay rate. The galactic contribution is a complicated function \eqref{pb34} of the kinematic parameters of the Solar System and the binary pulsar, which are available from fundamental catalogs.  Current determinations of the kinematic parameters entering the right hand-side of \eqref{pb34} 
are so reliable that the galactic contribution can be accurately computed and removed from the right hand-side of \eqref{pb23}. The remaining deviation of the predicted value of the orbital decay rate from its observed value is
\ba
\le(\frac{\dot P_b}{P_b}\ri)^{\rm GR}_{\rm pred}-\le(\frac{\dot P_b}{P_b}\ri)^{\rm GR}=0.9983\pm 0.0016\;,
\ea 
i.e., the theory is validated within $0.16\%$ \citep{weisberg_2010ApJ,weisberg_2016ApJ}.

It is remarkable that this type of testing of general relativity can be inverted. As recently emphasized by \citet{Chakrabarti_2020}, assuming
that general relativity holds and, therefore, the gravitational-wave orbital decay contribution can be predicted and accurately subtracted,
the observed orbital period changes can be converted into measurements of the relative Galactic acceleration at the binary pulsar's and Sun's locations.
\citet{Chakrabarti_2020} used such relative accelerations derived from observations of 14 binary pulsars to constrain the local gravitational field produced by the Galactic potential.
They found the {\it Oort limit}, that is the total volume mass density in the Galactic mid-plane, to be equal to $0.08^{0.05}_{-0.02}\;{\rm M}_\odot\,{\rm pc}^{-3}$. Accounting for the baryonic mass budget derived in \citep{McKee_2015}, \citet{Chakrabarti_2020} obtained a local dark matter density equals to  $\rho_{\rm DM}=-0.004^{0.05}_{-0.02}\;{\rm M}_\odot\,{\rm pc}^{-3}$, which is lower than that obtained by \citet{McKee_2015} but consistent within the current standard estimates of its measurement in the solar neighborhood by independent methods \citep{Bovy_2012}.

\subsection{Gravitational waves}

Gravitational waves are the solutions of Einstein's equations describing the propagation of ripple perturbations of spacetime curvature with the speed of light.  In fact, the speed of gravitational waves is set to be equal to the speed of light by the Einstein's postulate that the fundamental speed of gravity $c_g$ appearing in the gravity sector of general relativity \footnote{ The gravity sector of general relativity is associated with gravitational variables entering the left-hand side of Einstein's field equations while the electromagnetic/matter sector is associated with the energy-momentum tensor of matter on the right-hand side of these equations. The speed of gravity is, therefore, the speed $c=c_g$ normalizing the time derivatives of the metric tensor, $g_{\m\n}$, which are a constitutional part of both the Christoffel symbols and the Riemann tensor \citep{kopfoma_2006FoPh,kopfom_2007GReG}.} has the same numerical value as the speed of light $c$ from the electromagnetic/matter sector \citep{Kopeikin_2004CQGra,kopeikin_2006IJMPD}. This postulate has been recently confirmed experimentally by measuring the bending of light by the moving gravitational field of Jupiter \citep{kop_2001ApJ,Fomalont_2003ApJ} and in observations of gravitational waves \citep{Cornish_2017PhRvL}, in complete agreement with general relativity.

Gravitational waves were detected in 2015 by the Laser Interferometry Gravitational-wave Observatory (LIGO) \citep{Reitze_2017PhyU,Schutz_2018RSPTA}. 
 This discovery opened a new window to study the nature of black holes, neutron stars and astrophysical processes in the very early universe, which are not accessible for observations by any other means \citep{Schutz_1999CQG}. The gravitational-wave spectrum is distributed approximately over 10 decades of frequency, from the high-frequency normal modes of oscillating neutron stars down to the lowest frequencies produced by various cosmological mechanisms such as inflation, phase transitions, cosmic strings, etc. \citep{allen_1997rggr,maggiore_2000PhR}. Plane gravitational waves produce periodic fluctuations in the apparent positions of distant astronomical sources and in the time of propagation of electromagnetic signals from them. These fluctuations have a characteristic amplitude that is proportional to the strain of the gravitational wave \citep{bragin_1990NCimB,Schutz_2010IAU}. Direct astrometric measurements of gravitational waves from a single source are not possible with the current pulsar timing or VLBI technology. However, it was noticed that the stochastic ensemble of these waves produced by various mechanisms can be detected in a roundabout way by studying temporal and angular correlations in the times of arrival (TOA) of radio pulses from the pulsar timing array (PTA) \citep{hellings_1983ApJ,foster_1990ApJ,levin_2009MNRAS,manchester_2013CQGra,babak_2017PhRvD}, or in the redshifts of quasars \citep{seto_2006PhRvD}, or in the pattern of their proper motions in an astrometric fundamental catalog \citep{gwinn_1996ApJ,gwinn_1997ApJ,kaiser_1997ApJ,jaffe_2004NewAR,makarov_2010IAUS,Flanagan_2011PRD,bini_2018PhRvD,Qin_2019PhRvD}.

The temporal two-point cross-correlation function in the times of arrival of radio signals is known as the Hellings-Downs curve \citep{hellings_1983ApJ,lee_2008ApJ}
\ba \label{op2}
C_{ab}&=&\frac{1}{12\pi^2}\zeta^{\rm HD}_{ab}\int_0^\infty|h_c(f)|^2f^{-3}df\;,\\\label{opr3}
\zeta^{\rm HD}_{ab}&=&\sin^2\frac{\theta_{ab}}{2}\le[3\ln\le(\sin\frac{\theta_{ab}}{2}\ri)-\frac14\ri]+\frac12\le(1-\delta_{ab}\ri)\;,
\ea
where $h_c(f)$ is the characteristic strain of the gravitational wave metric perturbation at frequency $f$, $\theta_{ab}$ is the angle between two sources of radio waves labeled $a$ and $b$ respectively, $\delta_{ab}$ is the Kronecker symbol. The magnitude of the strain at frequency $f$ is related to the dimensional energy density of gravitational waves $\Omega_g(f)$ at this frequency by
\ba
|h_c(f)|^2&=&\frac32\frac{H_0\Omega_g(f)}{\pi^2f^2}\;,
\ea
where $H_0$ is the present value of the Hubble constant.

The shape of the Helling-Downs curve \eqref{opr3} provides a  means to measure polarization properties of gravitational waves and to detect possible violations of general relativity \citep{lee_2008ApJ,nano-hertz-gw_2019A&A}. Measuring the amplitude of the temporal cross-correlation function \eqref{op2} allows us to evaluate the overall energy density of the ensemble of gravitational waves as well as to get information about their spectrum. The PTA technique is the most sensitive to gravitational waves in the frequency band ranging from 1 nanohertz to a few tens of microhertz, where the mergers of supermassive black hole binaries (SMBHBs) are considered to be the most promising sources of gravitational waves. Preliminary results on the spectrum, energy and other characteristics of gravitational waves emitted by SMBHBs have been recently obtained from analysis of the 11-year data release of monitoring of 45 millisecond pulsars by the North American Nanohertz Observatory for Gravitational Waves (NANOGrav) \citep{nanograv_2018ApJS} and published in \citep{nanograv_2019ApJ,nanograv_2020ApJ_1,nanograv_2020ApJ_2}.       

The angular two-point cross-correlation function has been derived by \citet{gwinn_1996ApJ,gwinn_1997ApJ}, who focused mainly on the contribution of the second spherical harmonic in the expansion of the proper motion of light sources 
caused by stochastic gravitational waves over the entire sky. Recently, these calculations have been extended to all spherical harmonics by \citet{Flanagan_2011PRD} and \citet{Klioner_2018CQG} and we summarize them below for the reader's convenience. Let the source of light be observed in the direction defined by a unit vector ${\bm n}=(n^i)=\le(\cos\a\cos\d,\sin\a\cos\d,\sin\d\ri)$ where the angle $\a$ is the right ascension  and $\d$ is the declination. By definition, the proper motion is a time derivative of the unit vector, ${\bm\mu}=\dot{\bm n}$ which is a vector field on a unit sphere. The angular cross-correlation function of proper motions of distant sources caused by a stochastic gravitational wave ensemble, is 
\ba
C^{ij}_{ab}&=&\langle\mu^i_a\mu^j_b\rangle\;,
\ea
where the angular brackets denote the averaging over the ensemble of stochastic gravitational waves, the (geometric) vector indices $i$ and $j$ take values $\{1,2,3\}$, the labels $a$ and $b$ numerate the sources from the astrometric catalog seen in the directions of unit vectors ${\bm n}_a=(n_a^i)$ and ${\bm n}_b=(n_b^i)$ respectively, and $\mu^i_a=\dot{n}^i_a$, $\mu^i_b=\dot{n}^i_b$.
The Book-Flanagan cross-correlation function reads \citep{Flanagan_2011PRD}
\ba\label{juq2}
C^{ij}_{ab}&=&H_0^2\zeta^{\rm BF}_{ab}\le(A^iA^j+B^iC^j\ri)\int_0^\infty\Omega_g(f)d\ln f\;,\\
\zeta^{\rm BF}_{ab}&=&\frac12-6\sin^2\le(\frac{\theta_{ab}}{2}\ri)\le[\tan^2\le(\frac{\theta_{ab}}{2}\ri)\ln\le(\sin\frac{\theta_{ab}}{2}\ri)+\frac{7}{12}\ri]\;,
\ea
where $\theta_{ab}$ is the angle between the two sources, that is, $\cos\theta_{ab}={\bm n}_a\cdot{\bm n}_b$, and three mutually-orthogonal vectors
\be
{\bm A}=\sin^{-1}\theta_{ab}\le({\bm n}_a\times{\bm n}_b\ri)\qquad,\qquad {\bm B}={\bm n}_a\times{\bm A}\qquad,\qquad {\bm C}={\bm n}_b\times{\bm A}\;.
\ee

The vector field of proper motions caused by stochastic gravitational waves can be decomposed with respect to the 
 basis of vector spherical harmonics
\ba
{\bm\mu}&=&\sum_{l=2}^\infty\sum_{m=-l}^{+l}\le[\mu_{Elm}{\bm Y}^E_{lm}({\bm n})+\mu_{Blm}{\bm Y}^B_{lm}({\bm n})\ri]\;,
\ea
where ${\bm Y}^E_{lm}({\bm n})$ and ${\bm Y}^B_{lm}({\bm n})$ are the electric- and magnetic-type transverse vector spherical harmonics \citep{thorne_1980}. The cross-correlation matrix of the angular spectral decomposition is 
\ba
\langle \mu_{Qlm}\mu_{Q'l'm'}\rangle&=&H_0^2 C^{QQ'}_{lml'm'}\int_0^\infty\Omega_g(f)d\ln f\;,
\ea
where the labels $Q\in\{E,B\}$ and $Q'\in\{E,B\}$. The coefficients \citep{Flanagan_2011PRD}
\ba\label{iod4}
C^{QQ'}_{lml'm'}&=&\oint\oint d^2\Omega_{{\bm n}_a}d^2\Omega_{{\bm n}_b}Y^*_{lm}({\bm n}_a)Y_{l'm'}({\bm n}_b)\nabla^a_i\nabla^b_j\le[\zeta^{\rm BF}_{ab}\le(A^iA^j+B^iC^j\ri)\ri]\;,
\ea
where $Y_{lm}({\bm n}_a)$ are the scalar spherical harmonics, the integration is over the unit sphere, $\nabla^a_i$ and $\nabla^b_i$ are the normalized 3-dimensional gradients with respect to $x^i_a$ and $x^i_b$ entering the definition of the unit vectors: $n^i_a=x^i_a/|{\bm x}_a|$ and $n^i_b=x^i_b/|{\bm x}_b|$.
Computation by \eqref{iod4} yields \citep{Flanagan_2011PRD}
\ba
C^{QQ'}_{lml'm'}&=&\frac12\d_{QQ'}\d_{ll'}\d_{mm'}\frac{4\pi}{2l+1}\a_l\;,
\ea
where
\ba
\alpha_l&=&24(2l+1)\frac{(l-1)!}{(l+1)!}\frac{(l-2)!}{(l+2)!}\;.
\ea

Statistical power of the expectation value of the proper motion induced by the stochastic gravitational waves is
\ba
\langle{\bm\mu}^2\rangle&=&\sum_{Qlm}\sum_{Q'l'm'}{\bm Y}^Q_{lm}({\bm n}){\bm Y}^{Q'}_{l'm'}({\bm n})\langle \mu_{Qlm}\mu_{Q'l'm'}\rangle=\sum_{l=2}^\infty\langle\mu_l^2\rangle\;,
\ea
where
\ba
\langle\mu_l^2\rangle&=&\alpha_l H_0^2\int_0^\infty \Omega_g(f)d\ln f\;
\ea
and the numerical coefficients $\alpha_l$ characterize the power of the $l$-th spherical harmonic in the proper motion spectrum. \citet{gwinn_1997ApJ} calculated $\alpha_2=5/6$. \citet{Flanagan_2011PRD} provided numerical estimates for next 10 coefficients up to $l=11$. \citet{Klioner_2018CQG}  derived a general mathematical expression for all harmonics,
\ba
\alpha_l&=&24(2l+1)\frac{(l-1)!}{(l+1)!}\frac{(l-2)!}{(l+2)!}\;.
\ea
The expansion coefficients $\a_l$ decrease asymptotically as $l^{-5}$ as the harmonic degree $l$ increases.

The formalism of the astrometric detection of ultra-long gravitational waves described above can be used in future astrometric missions having much better precision than Gaia \citep{Kopeikin_2000tmcs}. The current level of astrometric technology does not allow us to set useful physical constraints of the density of primordial gravitational waves in the early universe \citep{Schutz_2010IAU}, although such an attempt has been done with radio-quasars \citep{gwinn_1996ApJ,gwinn_1997ApJ}. The main reasons for the insufficient sensitivity of astrometric detection of gravitational waves are relatively large accidental and systematic errors of the coefficients of vector spherical harmonics representing the observed proper motion field of CRF-defining quasars and AGNs as well as the occurrence of post-fit residuals with perturbations (outliers) beyond the statistical expectation. The latter emerges in comparison analyses of Gaia and ICRF astrometry \citep{Makarov_2017ApJ}, and may partly be caused by the morphology of the optical counterparts. Users of the radio-optical CRF should also remember that the global spin of the Gaia celestial coordinate system has been technically adjusted to zero on the specially selected sample of reference objects. 

\section{Future of Fundamental Catalogs}

There are two basic directions in the progressive development of astrometric catalogs establishing the inertial reference frame for various applications in astronomy and fundamental physics. They are concerned with the continuous improvement of astrometric precision in positions, proper motions, and parallaxes, and the increase in the number of catalog's objects. Currently, the astrometric precision of fundamental radio and optical catalogs is about 100 $\mu$as \citep{ICRF3_2018evn,Gaia-CRF2,icrf3_2020A&A}. It was recognized long ago that at this level of precision, the data processing of observations used for the construction of catalogs would require an elaborate general-relativistic model of celestial and terrestrial reference frames \citep{iau2000,kopeikin_2011book,Soffel_2013book} and propagation of light  \citep{klikop_1992AJ,klioner_2003AJ}, along with a substantially accurate description of motion of the Solar System bodies  \citep{pitjeva_2001CeMDA,standish_2002A&A,pitjeva_2005SoSyR,fienga_2008A&A} and the observer (Earth rotation parameters) \citep{capitaine_2009CeMDA,capitaine_2012RAA,dehant_mathews_2015}. These relativistic models have been worked out by a number of research groups around the globe and summarized in the form of standards of the International Earth Rotation and Reference Systems Service (IERS) \citep{IERS_2010conventions}. 

The near-future goal of fundamental astronomy is to attain the precision of 1 $\mu$as \citep{fomalont_2004NewAR,reid_2014ARA&A,theia_2016SPIE,lattanzi_2018SPIE,rioja_2019arXiv} and to progress beyond this threshold as soon as technology permits \citep{lindegren_2007HiA,zschocke_2014jsrs,brown_2014EAS}. In a more distant future, it may be possible to achieve a sub-microarcsecond astrometric resolution \citep{iau_co180_2000}. Data processing of sub-microarcsecond astrometric observations will require taking into account a great deal of relativistic effects both in the motion of the observer and the source of light as well as in the propagation of light rays \citep{Kopeikin_2000tmcs}. Relativistic models of equations of motion of the Solar System will have to be extended, at least, to the level of the 2-nd post-Newtonian approximation and to incorporate a large number of gravitational relativistic multipoles of the Solar System's bodies \citep{racine_2005PhRvD,kop_2019EPJP,k2019PRD} while the propagation of light must include various relativistic effects in the time delay and deflection angle caused by the higher-order post-Newtonian terms \citep{teyssandier_2012CQGra,teyssandier_2013CQGra} along with the orbital and rotational motion of the light-ray deflecting bodies \citep{kopeikin_1999PhRvD-sch,kopeikin_2002PhRvD_mash,kopeikin_2009MNRAS} and their complicated multipolar structure \citep{kopeikin_1999PhRvD,koppolkor2006,kopmak_2007PhRvD}.

Beyond the sub-microarcsecond threshold, one will see a new population of celestial physical phenomena caused by the presence of primordial gravitational waves from the early universe and different localized astronomical sources like binary stars, space-time topological defects (cosmic strings), moving gravitational lenses, the time variability of gravitational fields of the Solar System and the sources of light, and many others -- see the review by \citet{vallenari_2018FrASS} for more detail. Adequate physical interpretation of these yet undetectable sub-microarcsecond phenomena cannot be achieved  within the currently applied models, which will require significant development and extension into the field of relativistic gravitational physics \citep{klioner_2012MmSAI} and comprehensive understanding of the astrophysical origin of the astrometric jitter in position, proper motion, and structure of the observed light sources \citep{vallenari_2018FrASS}. 

A practical realization of sub-microarcsecond fundamental catalogs can be achieved with a number of different astronomical techniques in various ranges of the electromagnetic spectrum. In particular, radio astrometry at this level of accuracy adhering to the time-tested concept of absolute measurements requires technical facilities like VLBA {\color{blue}[\url{https://science.nrao.edu/facilities/vlba/docs/manuals/oss}]}, VERA \cite{vera_2020_performance} and the Square Kilometer Array (SKA)  \citep{fomalont_2004NewAR,godfrey_2012PASA,reid_2014ARA&A,SKA_2020PASA} with baselines ranging from the Earth's radius to the size of the lunar orbit. First successful steps in this direction have been recently taken with the space radio-interferometric mission RadioAstron \citep{kardashev_7910147,popov_8738406,gurvits_2020} which proved to be very productive {\color{blue}[\url{http://www.asc.rssi.ru/radioastron/publications/publ.html}]}.

Looking into the future of precision optical astrometry, the prospects seem to be less well defined, because it is not obvious which technological innovation would bring a significant jump in accuracy and productivity compared to the currently operating Gaia mission. The Space Interferometry Mission (SIM) was a bold initiative to move beyond 1 $\mu$as  in differential astrometry by replacing a traditional telescope with an imaging camera with a Michelson-type optical interferometer \citep{2008PASP..120...38U}. The theoretical astrometric precision of phase referenced interferometers is defined by the ratio $\lambda/B$, where $\lambda$ is the effective wavelength and $B$ is the baseline length, for photon shot noise limited, monochromatic interferometric fringes. The corresponding performance limit of a traditional imaging telescope is proportional
to $\lambda/D$, with $D$ being the entrance pupil diameter. Thus, interferometry can in principle provide much better astrometric precision by using baselines exceeding the largest feasible collecting areas of space telescopes. SIM, however,
was much limited by budgetary constraints and the basic technical design of a single-body space craft. 

The astrometric capabilities of optical interferometry could be indefinitely improved by operating a set of siderostats in a formation-flying
mode with a separate beam combiner space craft, similar to the now defunct DARWIN concept \citep{2004AdSpR..34..613F}, but
without the nulling function for direct imaging. The technical challenges of monitoring and maintaining long baselines between
free-floating telescopes within microns are formidable, however. In the domain of proven technology, the Theia initiative
\citep{theia_2017arXiv} promises to surpass the 1 $\mu$as threshold for brighter targets, but only in the narrow-angle
regime of differential astrometry, which can hardly contribute to fundamental astrometry. The proposed Gaia-NIR project
\citep{2019arXiv190712535H},
a successor to Gaia in the near infrared, will be truly a global astrometry mission greatly improving the Celestial Reference Frame by obtaining a second epoch of reference object positions, observing deeper into the dusty Galactic belt and beyond, and measuring fainter infrared quasars and AGNs, but only a moderate progress in the single measurement accuracy is anticipated. 

\section*{Acknowledgments} 
We thank Dr. Brian Luzum (United States Naval Observatory, USA), Prof. L\'aszl\'o Szabados (Konkoly Observatory, Hungary) and the referees for valuable comments and suggestions which helped us to improve presentation of the manuscript. We are grateful to Dr. Yuri Y. Kovalev (ASC Lebedev Physical Institute, Russia) and Dr. Jo Bovy (University of Toronto, Canada) for helpful discussions and useful bibliographic references. 
 
\bibliographystyle{unsrtnat}
\bibliography{Fundamental_Astronomy-references_final}

\end{document}